\documentclass[preprint,12pt]{elsarticle}
\usepackage{amssymb}

\usepackage[ruled,noline]{algorithm2e}
\usepackage{extarrows}
\usepackage{amsmath}
\usepackage{mathrsfs}
\usepackage{bbm}
\usepackage{booktabs}
\usepackage{subfigure}
\makeatletter
\def\ps@pprintTitle{%
  \let\@oddhead\@empty
  \let\@evenhead\@empty
  \def\@oddfoot{\reset@font\hfil\thepage\hfil}
  \let\@evenfoot\@oddfoot
}
\makeatother

\makeatletter

\newcommand {\Rmnum} [1] {\expandafter \@slowromancap \romannumeral #1@}
\makeatother
\newtheorem{Conjecture}{Conjecture}
\newtheorem{Theorem}{Theorem}
\newtheorem{Definition}{Definition}

\newtheorem{Lemma}{Lemma}

\begin{document}

\begin{frontmatter}

\title{Indistinguishability and semantic security for quantum encryption scheme}


\author[author1]{Chong Xiang}
\author[author1]{Li Yang\corref{cor1}}

\cortext[cor1]{Corresponding author. E-mail: yangli@iie.ac.cn}
\address[author1]{State Key Laboratory of Information Security, Institute of Information Engineering, Chinese Academy of Sciences, Beijing 100093, China}
\begin{abstract}
We investigate the definition of security for encryption scheme in quantum context. We systematically define the indistinguishability and semantic security for quantum public-key and private-key encryption schemes, and for computational security, physical security and information-theoretic security. Based on our definition, we present a necessary and sufficient condition that leads to information-theoretic indistinguishability for quantum encryption scheme. The equivalence between the indistinguishability and semantic security of quantum encryption scheme is also proved.

\end{abstract}


\end{frontmatter}


\section{Introduction}
The definition of security for encryption scheme is an important area of cryptography. Up till now, both the quantum public-key encryption \cite{Ben84,Yan03,Nik08,Gao08,Yan1012} and quantum private-key encryption \cite{Boy03,Amb00,Amb04} has been carried out.  Here we investigate the indistinguishability and semantic security into quantum context which would be useful for analysis the security of quantum encryption schemes.

모모In our previous work, we have already shown the definition of the indistinguishability for quantum public-key encryption scheme\cite{Pan10}, for quantum private-key encryption scheme\cite{Cho12},and for quantum bit commitment scheme and have presented a necessary and sufficient condition leads to this security\cite{Cho10}. Here we will systematically define the indistinguishability and semantic security for quantum public-key and private-key encryption schemes, and for computational security, physical security and information-theoretic security.

모모The quantum parameters are continuous variable. In order to give the definition of indistinguishability for quantum encryption scheme, we first give a definition of indistinguishability for encryption scheme with continuous variable based on probability density function, and a definition of indistinguishability based on multi-circuits. We show that these definitions are equivalent. Then we prove that the indistinguishability based on multi-circuits is equivalent to ordinary indistinguishability with single-circuit. Then we get the definition of indistinguishability for quantum encryption scheme. Similarly we define the semantic security.

모모The equivalence between computational indistinguishability and semantic security for classical encryption scheme is already proved, but the equivalence for information-theoretic ones is still an open problem. For public-key encryption scheme, there is no information-theoretically secure classical public-key encryption scheme, so we discuss the equivalence between computational indistinguishability and semantic security for quantum encryption scheme and between information-theoretic ones. About private-key encryption scheme, the equivalence between computational indistinguishability and semantic security for quantum encryption scheme and between information-theoretic ones for classical and quantum encryption schemes all are discussed.

\section{Preliminaries}
The definitions of indistinguishability and semantic security were firstly presented by S. Goldwasser and S. Milcali\cite{God84,God82}, then Goldrich\cite{Gold04}
developed these definitions and classify defined the indistinguishability and semantic security with different conditions.

\subsection{Indistinguishability}
The indistinguishability for private-key encryption scheme is:
\begin{Definition}\label{CIP}{\em(indistinguishability for private-key encryption scheme):} A private-key encryption scheme, $(G, E, D)$, is said to be an indistinguishable scheme if for every polynomial-size circuit family $\{C_n\}$, every positive polynomial $p(\cdot)$, all sufficiently large $n$, and every $x,y\in\{0,1\}^{Poly(n)}$,
\end{Definition}\begin{eqnarray}
  |\textrm{Pr}[C_n(E_{G_1(1^n)}(x))=1]- \textrm{Pr}[C_n(E_{G_1(1^n)}(y))=1]|<\frac{1}{p(n)}.
\end{eqnarray}

For public-key encryption scheme, the indistinguishability is defined as:
\begin{Definition}\label{CIPU}{\em(indistinguishability for public-key encryption scheme):} A public-key encryption scheme, $(G, E, D)$, is said to be an indistinguishable scheme if for every polynomial-size circuit family $\{C_n\}$, every positive polynomial $p(\cdot)$, all sufficiently large $n$, and every $x,y\in\{0,1\}^{Poly(n)}$,
\end{Definition}\begin{eqnarray}
  |\textrm{Pr}[C_n(G_1(1^n), E_{G_1(1^n)}(x))=1]- \textrm{Pr}[C_n(G_1(1^n), E_{G_1(1^n)}(y))=1]|<\frac{1}{p(n)}.
\end{eqnarray}

These definitions are based on computational security, if the inequalities are satisfied for every circuit family $\{C_n\}$ instead of for every polynomial-size circuit family $\{C_n\}$, we gains the definitions based on information-theoretic security.

\subsection{Semantic security}
The semantic security for private-key encryption scheme is shown as:
\begin{Definition}\label{CSP}{\em(semantic security for private-key encryption scheme):}

A private-key encryption scheme, $(G, E, D)$, is said to be semantically secure if for every probabilistic polynomial-time algorithm $A$ there exists a probabilistic polynomial-time algorithm $A'$ such that for every probability ensemble $\{X_n\}_{n\in\mathbb{N}}$, with $|X_n|\leq poly(n)$, every pair of polynomially bounded functions $f(\cdot),h(\cdot)$: $\{0,1\}^*\rightarrow \{0,1\}^*$, every positive polynomial $p(\cdot)$ and all sufficiently large $n$,
\begin{eqnarray}
  &&\mathrm{Pr}[A(1^n, E_{G_1(1^n)}(X_n), 1^{|X_n|}, h(1^n, X_n))=f(1^n, X_n)]\nonumber\\
  &<&\mathrm{Pr}[A'(1^n, 1^{|X_n|}, h(1^n, X_n))=f(1^n, X_n)+\frac{1}{p(n)}.
\end{eqnarray}
\end{Definition}

For public-key encryption scheme, it is:
\begin{Definition}\label{CSPU}{\em(indistinguishability for public-key encryption scheme):}

A public-key encryption scheme, $(G, E, D)$, is said to be semantically secure if for every probabilistic polynomial-time algorithm $A$ there exists a probabilistic polynomial-time algorithm $A'$ such that for every $\{X_n\}_{n\in\mathbb{N}}$,$f(\cdot),h(\cdot)$밃$p(\cdot)$ and $n$ as in Definition \ref{CSP},
\end{Definition}\begin{eqnarray}
  &&\textrm{Pr}[A(1^n, G_1(1^n), E_{G_1(1^n)}(X_n), 1^{|X_n|}, h(1^n, X_n))=f(1^n, X_n)]\nonumber\\
  &<&\textrm{Pr}[A'(1^n, 1^{|X_n|}, h(1^n, X_n))\nonumber\\
  &=&f(1^n, X_n)+\frac{1}{p(n)}.
\end{eqnarray}

Similarly, These definitions are based on computational security, if the inequalities are satisfied that for every algorithm $A$ there exists a probabilistic algorithm $A'$ instead of for every polynomial-time algorithm $A$ there exists a probabilistic polynomial-time algorithm $A'$, we gains the definitions based on information-theoretic security.

\section{Indistinguishability for quantum encryption scheme}
Firstly, we discuss the indistinguishability of quantum private-key encryption scheme based on that of classical private-key encryption scheme.

\subsection{Indistinguishability based on probability density function}
As the quantum information is continuous character, if we want to define the indistinguishability of quantum encryption scheme, firstly we should present the indistinguishability of continuous variable. It must depend on the probability density function, so we give the C-indistinguishability of classical information as follow:
\begin{Definition}\label{CIPC}
If the plaintext is continuous variable, let the probability density function of plaintext space $P$ is $q(x)$, which is a continuous function. A private-key encryption scheme, $(G, E, D)$, is said to be a C-indistinguishable scheme if it satisfies the condition as follow: for every polynomial-size circuit families $\{C_n\}$, every positive polynomial $p(\cdot)$, all sufficiently large $n$, and every $x,y\in P$,
\begin{eqnarray}
  \left|\mathrm{Pr}[C_n(E_{G_1(1^n)}(x))=1]-\mathrm{Pr}[C_n(E_{G_1(1^n)}(y))=1]\right|<\frac{1}{p(n)}.
\end{eqnarray}
\end{Definition}

\subsection{Indistinguishability based on multi-circuits}
Then we show a definition of indistinguishability based on multi-circuits as follow:

\begin{Definition}\label{CIPM}
A private-key encryption scheme, $(G, E, D)$, is said to be a M-indistinguishable scheme if for every polynomial-size circuit families $\{C_n^i\}$, here $i=1,2\cdots,m$, every positive polynomial $p_i(\cdot)$, all sufficiently large $n$, and every $x_i,y_i\in\{0,1\}^{Poly(n)}$,
\begin{eqnarray}
  |\mathrm{Pr}[C_n^1(E_{G_1(1^n)}(x_1))=1]&-& \mathrm{Pr}[C_n^1(E_{G_1(1^n)}(y_1))=1]|<\frac{1}{p_1(n)},\nonumber\\
  |\mathrm{Pr}[C_n^2(E_{G_1(1^n)}(x_2))=1]&-& \mathrm{Pr}[C_n^2(E_{G_1(1^n)}(y_2))=1]|<\frac{1}{p_2(n)},\nonumber\\
  &\vdots&\nonumber\\
  |\mathrm{Pr}[C_n^m(E_{G_1(1^n)}(x_m))=1]&-& \mathrm{Pr}[C_n^m(E_{G_1(1^n)}(y_m))=1]|<\frac{1}{p_m(n)}.
\end{eqnarray}
\end{Definition}

\subsection{Equivalence of the definitions}
Based on above definitions of indistinguishability, we will prove that they are all equivalence. The proofs in this section are all based on definitions of computational security.

\begin{Lemma}\label{PTM}
If a private-key encryption scheme is said to be a M-indistinguishable scheme if and only if it is an indistinguishable scheme.
\end{Lemma}

{\bf Proof.} For both sufficiency and necessity, we can prove with reduction to absurdity. Here we prove the sufficiency for example:

If the scheme is not M-indistinguishable, there must exist at least a polynomial-size circuit family $\{C_n^i\}$, a positive polynomial $p_i(\cdot)$, and $x_i,y_i\in\{0,1\}^{Poly(n)}$, which lead to that for all sufficiently large $n$
    \begin{eqnarray}
      |\textrm{Pr}[C_n^i(E_{G_1(1^n)}(x_i))=1]- \textrm{Pr}[C_n^i(E_{G_1(1^n)}(y_i))=1]|\geq\frac{1}{p_i(n)}.
    \end{eqnarray}
  Therefore, if we let $\{C_n\}=\{C_n^i\}$, $p(\cdot)=p_i(\cdot)$, $x=x_i,y=y_i$, we can get:
  \begin{eqnarray}
    |\textrm{Pr}[C_n(E_{G_1(1^n)}(x))=1]- \textrm{Pr}[C_n(E_{G_1(1^n)}(y))=1]|\geq\frac{1}{p(n)},
  \end{eqnarray}
  for all sufficiently large $n$, which means this scheme is not indistinguishable.

  Thus the sufficiency is proved.

  Similarly, we can use  prove the necessity. $\Box$

\begin{Lemma}\label{PTC}
If a private-key encryption scheme with continuous plaintext is said to be a C-indistinguishable scheme if it is a indistinguishable scheme.
\end{Lemma}

{\bf Proof.} Assume the scheme is not a C-indistinguishable scheme, there must exist $x,y\in P$, which satisfy that for all sufficiently large $n$, every polynomial-size circuit families $\{C_n\}$, every positive polynomial $p(\cdot)$:
\begin{eqnarray}
  \left|\mathrm{Pr}[C_n(E_{G_1(1^n)}(x))=1]-\mathrm{Pr}[C_n(E_{G_1(1^n)}(y))=1]\right|\geq\frac{1}{p(n)}.
\end{eqnarray}
Let $n_0=max\{|x|,|y|\}$, here $|x|$ is the length of $x$, and let $n>n_0$, we can get that:
there exist $x,y\in\{0,1\}^{Poly(n)}$, which satisfy that for all sufficiently large $n$, every polynomial-size circuit families $\{C_n\}$, every positive polynomial $p(\cdot)$:
\begin{eqnarray}
  \left|\mathrm{Pr}[C_n(E_{G_1(1^n)}(x))=1]-\mathrm{Pr}[C_n(E_{G_1(1^n)}(y))=1]\right|\geq\frac{1}{p(n)}.
\end{eqnarray}

This reaches a contradiction to the hypothesis that the scheme is a indistinguishable scheme. Thus the lemma follows. $\Box$

The equivalence of these definitions is almost proved except the necessity of lemma.\ref{PTC}, we planed to complete this side via the definition based on multi-circuits, but it has not worked out yet, so it is still a conjecture.

\begin{Conjecture}\label{MTC}
If a private-key encryption scheme with continuous plaintext is said to be a C-indistinguishable scheme if and only if it is a M-indistinguishable scheme.
\end{Conjecture}

\subsection{Indistinguishability for quantum encryption scheme}
As the indistinguishability of classical private-key encryption scheme can lead to that of continuous variable, We suggest here a definition of information-theoretic indistinguishability for quantum private-key encryption scheme as follow:

\begin{Definition}\label{QIP}
A quantum private-key encryption scheme is information-theoretically indistinguishable if for every quantum circuit family \{$C_n$\}, every positive polynomial $p(\cdot)$, all sufficiently large $n$'s, and every $x,y\in\{0,1\}$:
\begin{eqnarray}
\Big|\mathrm{Pr}[C_n(E_{G(1^n)}(x))=1]-\mathrm{Pr}[C_n(E_{G(1^n)}(y))=1]\Big|<\frac{1}{p(n)},
\end{eqnarray}
where the encryption algorithm $E$ should be a quantum algorithm, and the ciphertext $E(x), E(y)$ are quantum states.
\end{Definition}

Similarly, for quantum public-key encryption scheme, the information-theoretic indistinguishability is shown as:

\begin{Definition}\label{QIPU}~~
A quantum public-key encryption scheme is information-theoretically indistinguishable if for every quantum circuit family \{$C_n$\}, every positive polynomial $p(\cdot)$, all sufficiently large $n$'s, and every $x, y$ in plaintext space:
\begin{eqnarray}
\Big|\mathrm{Pr}[C_n(G(1^n), E_{G(1^n)}(x)=1]-\mathrm{Pr}[C_n(G(1^n), E_{G(1^n)}(y)=1]\Big|<\frac{1}{p(n)},
\end{eqnarray}
where the encryption algorithm $E$ should be a quantum algorithm, and the ciphertext $E(x), E(y)$ are quantum states.
\end{Definition}

In classical context, the security is defined under two conditions, here the quantum definitions can be classified by three different conditions:
\begin{enumerate}
  \item As defined above, we get the definitions of information-theoretic indistinguishability.
  \item If the inequalities are satisfied for polynomial-size quantum circuit family $\{C_n\}$ instead of for every circuit family $\{C_n\}$, it results the definitions of computational indistinguishability.
  \item If the inequalities are satisfied for specifical exponential-size quantum circuit family $\{C_n\}$밃 it results the definitions of physical indistinguishability, here the size is determined by the protocol.
\end{enumerate}

 The physical security we presented here means that even it may be not information-theoretical secure, the way to attack is unable to realize limited to the objective physical conditions.

\subsection{The necessary and sufficient condition for information-theoretic indistinguishability}
Here we present the sufficient and necessary condition of the information-theoretic indistinguishability for quantum private-key encryption scheme as follow:

\begin{Theorem}\label{the1}
For every plaintexts $x$ and $y$ and key $k$, let the density operators of cipher states $\sum_kp_kE_k(x)$ and $\sum_kp_kE_k(y)$ are $\rho_x$ and $\rho_y$, respectively. A quantum private-key encryption scheme is said to be information-theoretically indistinguishable if for every positive polynomial $p(\cdot)$ and every sufficiently large  $n$,
\end{Theorem}\begin{eqnarray}
D(\rho_x,\rho_y)<\frac{1}{p(n)}.
\end{eqnarray}

{\bf Proof.}  For every quantum circuit family $\{C_n\}$,
\begin{eqnarray}
& &\mathrm{Pr}[C_n(E_{G(1^n)}(x))=1]\nonumber\\
&= &\sum_kp_k\cdot\mathrm{Pr}[C_n(E_k(x)\otimes \sigma)=1]\nonumber\\
&= &\mathrm{Pr}[C_n(\sum_kp_kE_k(x)\otimes \sigma)=1]\nonumber\\
&= &\mathrm{Pr}[C_n(\rho_x\otimes \sigma)=1],
\end{eqnarray}
where $\sigma$ is the density operator of service bits of $C_n$.

Similarly,
\begin{eqnarray}
\mathrm{Pr}[C_n(E_{G(1^n)}(y))=1]=\mathrm{Pr}[C_n(\rho_y\otimes \sigma)=1].
\end{eqnarray}

Any quantum circuit family ${C_n}$ built for distinguishing two density operators corresponds to a set of positive operator-values measure (POVM) \{$E_m$\}. Define $p_m=\mathrm{Tr}(C_n(\rho_x\otimes \sigma)E_m)$, $q_m=\mathrm{Tr}(C_n(\rho_y\otimes \sigma)E_m)$ the probabilities of measurement outcomes labeled by $m$. In this case, we have:
\begin{eqnarray}
& &\Big|\mathrm{Pr}[C_n(\rho_x\otimes \sigma)=1]-\mathrm{Pr}[C_n(\rho_y\otimes \sigma)=1]\Big|\nonumber\\
&\leq & \max_{\{E_m\}}\frac{1}{2}\sum_m|\mathrm{Tr}[E_m(C_n(\rho_x\otimes \sigma)-C_n(\rho_y\otimes \sigma))]\nonumber\\
&= & \max_{\{E_m\}}D(p_m, q_m).
\end{eqnarray}
The last formula is equal to
\begin{eqnarray}
D(C_n(\rho_x\otimes \sigma),C_n(\rho_y\otimes \sigma))\leq D(\rho_x\otimes \sigma,\rho_y\otimes \sigma)=D(\rho_x, \rho_y)<\frac{1}{p(n)} . \end{eqnarray}
Hence, according to the Definition \ref{QIP}, the theorem follows. $\Box$

For quantum public-key encryption scheme, we also have a theorem:
\begin{Theorem}\label{the1}
For every plaintexts $x$ and $y$ and public-key $k$, let the density operators of cipher states $\sum_kp_kE_k(x)$ and $\sum_kp_kE_k(y)$ are $\rho_x$ and $\rho_y$, respectively. A quantum private-key encryption scheme is said to be information-theoretically indistinguishable if for every positive polynomial $p(\cdot)$ and every sufficiently large  $n$,
\end{Theorem}\begin{eqnarray}
D(\rho_x,\rho_y)<\frac{1}{p(n)}.
\end{eqnarray}

the proof for quantum public-key encryption scheme is similar to the above.

\section{Semantic security for quantum encryption scheme}
The semantic security for quantum encryption scheme means that whatever can be efficiently computed from the ciphertext can be efficiently computed when given only the length of plaintext. For quantum private-key encryption scheme it turns out as:
\begin{Definition}\label{QSP}
A quantum private-key encryption scheme is semantically secure if for every quantum algorithm A there exist a quantum algorithm $A'$, such that for for every probability ensemble $\{X_n\}_{n\in\mathbb{N}}$, with $|X_n|\leq poly(n)$, every quantum bounded functions $f,h$, positive polynomial $p(\cdot)$, all sufficiently large $n$:
\begin{eqnarray}
&&\mathrm{Pr}[A(1^n, E_{G_1(1^n)}(X_n), 1^{|X_n|}, h(1^n, X_n))\nonumber\\
&=&f(1^n, X_n)]<\mathrm{Pr}[A'(1^n, 1^{|X_n|}, h(1^n, X_n))\nonumber\\
&=&f(1^n, X_n)]+\frac{1}{p(n)}.
\end{eqnarray}
where the encryption algorithm $E$ should be a quantum algorithm, and both algorithms and functions are output 0 or 1.
\end{Definition}

Note that here the probability function Pr include more parts than that within classical definitions, besides the probability distribution of $G$, $X_n$, $A$, $A'$, here as the quantum algorithms and functions are both output classical information, the function Pr should include the probability of collapse.

Similarly we can get the definition for quantum public-key encryption scheme:
\begin{Definition}\label{QSPU}
A quantum public-key encryption scheme, $(G, E, D)$, is said to be semantically secure if for every quantum algorithm $A$ there exists a quantum algorithm $A'$ such that for every $\{X_n\}_{n\in\mathbb{N}}$,$f(\cdot),h(\cdot)$밃$p(\cdot)$ and $n$ as in Definition \ref{QSP},
\begin{eqnarray}
  &&\mathrm{Pr}[A(1^n, G_1(1^n), E_{G_1(1^n)}(X_n), 1^{|X_n|}, h(1^n, X_n))=f(1^n, X_n)]\nonumber\\
  &<&\mathrm{Pr}[A'(1^n, 1^{|X_n|}, h(1^n, X_n))\nonumber\\
  &=&f(1^n, X_n)+\frac{1}{p(n)}.
\end{eqnarray}
where the encryption algorithm $E$ should be a quantum algorithm, and both algorithms and functions are output 0 or 1.
\end{Definition}

As aforementioned, the semantic security can also be classified by three different conditions:

\begin{enumerate}
  \item As defined above, we get the definitions of information-theoretic semantic security.
  \item If the inequalities are satisfied while $A$ and $A'$ are bounded with polynomial-time, it results the definitions of computational semantic security.
  \item If the inequalities are satisfied while $A$ and $A'$ are bounded with specifical exponential-time, it results the definitions of physical semantic security, here the size is determined by the protocol.
\end{enumerate}

\section{Equivalence of the security definitions}
Firstly, we state and prove the following theorem for quantum private-key encryption scheme with computational security. The similar results hold for quantum public-key encryption schemes and for quantum private-key encryption scheme with information-theoretic security.

\begin{Theorem}
  A quantum private-key encryption scheme is semantically secure if and only if it is indistinguishable.
\end{Theorem}

{\bf Proof.}
\begin{enumerate}
  \item "indistinguishability" implies "semantic security".

  Firstly, As the scheme is indistinguishable, for every $C_n,p(\cdot), x,n$ as in Def.\ref{QIP} and $y=1^{|x|}$, we can get the following inequality:
  \begin{eqnarray}\label{EqI-S1}
  \Big|\mathrm{Pr}[C_n(E_{G(1^n)}(x))=1]-
  \mathrm{Pr}[C_n(E_{G(1^n)}(1^{|x|}))=1]\Big|<\frac{1}{p(n)},
  \end{eqnarray}

  Then we construct the quantum algorithm $A'$ as follow:
  The quantum algorithm $A'$ performs essentially while replace the input $X_n$ of algorithm $A$ with $1^{|X_n|}$.

  To simplify the notations, let $h_n(x)\doteq h(1^n,x)$, $f_n(x)\doteq f(1^n,x)$, $A_n(x)\doteq A(1^n,x)$ and omit $1^{|X_n|}$ from the inputs given to $A$, then using the construction of $A'$ we get:
  \begin{eqnarray}
  & &\mathrm{Pr}[A(1^n, E_{G_1(1^n)}(X_n), 1^{|X_n|}, h(1^n, X_n))=f(1^n, X_n)]\nonumber\\
  &=&\mathrm{Pr}[A_n(E_{G_1(1^n)}(X_n),h_n(X_n))=f_n(X_n)];\nonumber\\
  & &\mathrm{Pr}[A'(1^n, 1^{|X_n|}, h(1^n, X_n))=f(1^n, X_n)]\nonumber\\
  &=&\mathrm{Pr}[A_n(E_{G_1(1^n)}(1^{|X_n|}),h_n(X_n))=f_n(X_n)];
  \end{eqnarray}

  For every string $x_n\in\{X_n\}$, the values $f_n(x_n), h_n(x_n)$ are fixed, then we construct a quantum circuit $C_n$ as follow: on input  $x_n$, the circuit $C_n$ invokes $A_n(E_{G_1(1^n)}(x_n),h_n(x_n))$ and outputs 1 while $A_n$ outputs $f_n(x_n)$, otherwise, $C_n$ outputs 0. This circuit is indeed of polynomial size because $f_n(x_n)$ and $g_n(x_n)$ are polynomial length and $A$ is a polynomial time quantum algorithm.

  Thus we can get:
  \begin{eqnarray}
  \mathrm{Pr}[C_n(E_{G(1^n)}(x))=1]
  =\mathrm{Pr}[A_n(E_{G_1(1^n)}(X_n),h_n(X_n))=f_n(X_n)];
  \end{eqnarray}

  Proof by contradiction, if the scheme is not semantically secure, then for every $A'$
  \begin{eqnarray}
    &&\mathrm{Pr}[A(1^n, E_{G_1(1^n)}(X_n), 1^{|X_n|}, h(1^n, X_n))=f(1^n, X_n)]\nonumber\\
    &>&\mathrm{Pr}[A'(1^n, 1^{|X_n|}, h(1^n, X_n))\nonumber\\
    &=&f(1^n, X_n)]+\frac{1}{p(n)},
  \end{eqnarray}
  which is equivalent to that:
    \begin{eqnarray}
    \mathrm{Pr}[C_n(E_{G(1^n)}(x))=1]-\mathrm{Pr}[C_n(E_{G(1^n)}(1^{|x|}))=1]>\frac{1}{p(n)},
  \end{eqnarray}
  this contradicts InEq.(\ref{EqI-S1}), so the sufficiency follows. $\Box$

  \item "semantic security" implies "indistinguishability".

  Also proof by contradiction, if the scheme is not indistinguishable, we can assume that there exists a polynomial $p(\cdot)$ and a polynomial-size circuit family $\{C_n\}$, such that for infinitely many n's there exist $x_n,y_n\in\{0,1\}^{poly(n)}$ so that:
  \begin{eqnarray}\label{EqI-S1}
  \Big|\mathrm{Pr}[C_n(E_{G(1^n)}(x_n))=1]-
  \mathrm{Pr}[C_n(E_{G(1^n)}(y_n))=1]\Big|>\frac{1}{p(n)},
  \end{eqnarray}

 ~~ Then we define $X_n$ is uniformly distributed over $\{x_n,y_n\}$, define $f(1^n,X_n)=1$ while $X_n=x_n$ and equals 0 while $X_n=y_n$ with both probability $1/2$, and define $h(1^n,X_n)$ equals the description of the circuit $C_n$ while it reveals no information on the value of $X_n$.

  Here we present a polynomial-time quantum algorithm $A$ that, it recovers $C_n=h(1^n,X_n)$, takes $E_{G(1^n)}(x_n)$ as input, and outputs what $C_n$ outputs.

  Thus we can get:
  \begin{eqnarray}\label{EQS-I1}
    & &\mathrm{Pr}[A(1^n, E_{G_1(1^n)}(X_n), 1^{|X_n|}, h(1^n, X_n))=f(1^n, X_n)]\nonumber\\
    &=&\frac{1}{2}\cdot\mathrm{Pr}[A(1^n, E_{G_1(1^n)}(x_n), 1^{|x_n|}, C_n)=1]\nonumber\\
    &&+\frac{1}{2}\cdot\mathrm{Pr}[A(1^n, E_{G_1(1^n)}(y_n), 1^{|y_n|}, C_n)=0]\nonumber\\
    &>&\frac{1}{2}+\frac{1}{2p(n)}
  \end{eqnarray}

  In contrast, while the input values $1^n,1^{|X_n|}$ and $h(1^n,X_n)$ are independent of the random variable $f(1^n,X_n)$, $A'$ can not output $f(1^n,X_n)$ with success probability above $1/2$, so we get:
  \begin{eqnarray}\label{EQS-I2}
    \mathrm{Pr}[A'(1^n, 1^{|X_n|}, h(1^n, X_n))=f(1^n, X_n)]\leq \frac{1}{2}.
  \end{eqnarray}

  Combining InEqs.(\ref{EQS-I1}),(\ref{EQS-I2}), we reach a contradiction to the hypothesis that the scheme is semantically secure. Thus the necessity follows. $\Box$

  As both sides of the theorem are proved, the theorem is proven. $\blacksquare$
\end{enumerate}

\section{Conclusions}
In this paper we suggest definitions of indistinguishability and sematic security for quantum encryption schemes with information-theoretic security, physical security and commotional security. We show that a necessary and sufficient condition leads to information-theoretic indistinguishability, which is useful for proving this security. We proved the equivalence between the indistinguishability and semantic security with computational security of quantum encryption schemes, the other equivalence is also hold with similar proof.

\section*{Acknowledgement}
This work was supported by the National Natural Science Foundation of China (Grant No. 61173157), Strategy Pilot Project of Chinese Academy of Sciences (Grant No. Sub-project XD06010702), and IIE몶s Cryptography Research Project.




\begin{thebibliography}{99}
\bibitem{Ben84}C. H. Bennett and G. Brassard, "Quantum cryptography: public key distribution and coin-tossing," Proceedings of IEEE International Conference on Computers, Systems and Signal Processing , pp. 175, 1984.
\bibitem{Yan03}L. Yang, "Quantum public-key cryptosystem based on classical NP-complete problem," e-print arXiv: quant-ph/0310076 , 2003.
\bibitem{Nik08}G. M. Nikolopoulos, "Applications of single-qubit rotations in quantum public-key cryptography," Phys. Rev. A 77, pp. 032348, 2008.
\bibitem{Gao08}F. Gao, Q. Y. Wen, S. J. Qin, and F. C. Zhu, "Quantum asymmetric cryptography with symmetric keys," Science in China Series G: Physics
Mechanics and Astronomy 52, pp. 1925, 2008.
\bibitem{Yan1012}L. Yang, M. Liang, B. Li, L. Hu, and D. G. Feng, "Quantum public-key cryptosystems based on induced trapdoor one-way transformations," e-
print arXiv: 1012.5249 , 2010.
\bibitem{Boy03}P. Boykin and V. Roychowdhury, "Optimal encryption of quantum bits," Phys. Rev. A 67(4), pp. 42317, 2003.
\bibitem{Amb00}A. Ambainis, M. Mosca, A. Tapp, and R. de Wolf, "Private quantum channel," Proc. 41st FOCS , pp. 547, 2000.
\bibitem{Amb04}A. Ambainis and A. Smith, "Small pseudo-random families of matrices: Derandomizing approximate quantum encryption," Proc. RANDOM, LNCS 3122, Berlin-Heidelberg-NewYork: Springer , pp. 249, 2004.
\bibitem{Pan10}J. Y. Pan and L. Yang, "Quantum public-key encryption with information theoretic security," e-print arXiv:1006.0354 , 2010.
\bibitem{Cho12}C. Xiang and L. Yang, "Quantum unicity distance," Proc. SPIE 8440, pp. 84400T, 2012.
\bibitem{Cho10}L. Yang, C. Xiang, and B. Li, "Qubit-string-based bit commitment protocols with physical security," e-print arXiv: 1011.5099 , 2010.
\bibitem{God84}S. Goldwasser and S. Micali, "Probabilistic encryption," Special issue of Journal of Computer and Systems Sciences 28(2), pp. 270, 1984.
\bibitem{God82}S. Goldwasser and S. Micali, "Probabilistic encryption and how to play mental poker keeping secret all partial information," Proceedings of the fourteenth annual ACM symposium on Theory of computing , pp. 365, 1982.
\bibitem{Gold04}O. Goldreich, "Foudations of cryptography: Basic applications," Cambridge University Press , 2001.
\end{thebibliography}
\end{document}